\documentstyle[preprint,aps]{revtex}
\tightenlines  
\begin{document} %\draft
\title{\LARGE \bf Ermakov Approach for
Minisuperspace Oscillators}

\author{H.C. Rosu and J. Socorro}
\address{Instituto de F\'{\i}sica IFUG, Apdo Postal E-143, Le\'on, Gto, Mexico}
\address{{\scriptsize May 29/99 Talk by H.C. Rosu at ICSSUR 6, Naples, Italy;
To be published in the Proceedings; gr-qc/9908028}}

%\author{Name of the second author}
%\address{Address of the second author}

\maketitle
\thispagestyle{empty}

\begin{abstract}
{\scriptsize
The WDW equation of arbitrary Hartle-Hawking factor ordering for several
minisuperspace universe models, such as the pure gravity FRW and Taub ones,
is mapped onto the
dynamics of corresponding classical oscillators. The latter ones are studied
by the classical Ermakov invariant method, which is a natural aproach in this
context.
For the more realistic case of a minimally coupled massive scalar field, one
can study, within the same type of approach, the corresponding squeezing
features as a possible means of describing cosmological evolution. Finally,
we comment on the analogy with the accelerator physics.}
\end{abstract}

\vspace{5mm}

{\bf 1}. The formalism of Ermakov-type invariants [1]
can be a useful, alternative method of investigating evolutionary and
chaotic dynamical problems in the ``quantum" cosmological
framework [2]. Moreover, the method
of adiabatic invariants is intimately related to geometrical angles and
phases [3] so that one may think of cosmological
Hannay's angles as well as various types of topological phases
as those of Berry and Pancharatnam [4].
In the following we apply the formal Ermakov scheme to some of the
simplest cosmological pure gravity oscillators, such as the empty
Friedmann-Robertson-Walker (EFRW) ``quantum" universes and the anisotropic
Taub ones.
The EFRW Wheeler-DeWitt (WDW)
minisuperspace equation reads [5]
%%%%%%%%%%%%%%
\begin{equation} \label{1}
\frac{d^2\Psi}{d\Omega^2}+Q\frac{d\Psi}{d\Omega}-\kappa e^{-4\Omega}\Psi
(\Omega) =0~,
\end{equation}
%%%%%%%%%%%%%%%
where $Q$, assumed nonzero [6] and kept as a free parameter, is the
Hartle-Hawking (HH) parameter for the factor ordering [7],
the variable $\Omega$ is Misner's time [8], and $\kappa$ is the curvature
index of the FRW universe; $\kappa =1,0,-1$ for closed, flat, open universes,
respectively.
For $\kappa=\pm 1$ the general solution is expressed in terms of
Bessel functions, $\Psi _{\alpha}(\Omega) =%\propto
e^{-2\alpha\Omega}\left(C_1I_{\alpha}(\frac{1}{2}e^{-2\Omega})+
C_2K_{\alpha}(\frac{1}{2}e^{-2\Omega})\right)$ and
$\Psi _{\alpha}(\Omega) =%\propto
e^{-2\alpha\Omega}\left(C_1J_{\alpha}(\frac{1}{2}e^{-2\Omega})+
C_2Y_{\alpha}(\frac{1}{2}e^{-2\Omega})\right)$, respectively,
where $\alpha =Q/4$.
The case $\kappa =0$ is special/degenerate, leads to hyperbolic functions
and will not be dealt with here.
%Using the change of function $\Psi=\Phi\exp(-Q\Omega/2)$ one can eliminate
%the first derivative and get
%%%%%%%%%%%%%%%
%\begin{equation} \label{1b}
%\Phi ^{''}-\left(\frac{Q^2}{4}+e^{-4\Omega}\right)\Phi=0
%\end{equation}
%%%%%%%%%%%%%%%%%%
Eq.~(1) can be mapped in a known way %[6]
to the canonical equations for a classical point
particle of %x=t,
mass $M=e^{Q\Omega}$, generalized coordinate $q=\Psi$,
momentum $p=
e^{Q\Omega}\dot{\Psi}$, (i.e., velocity $v=\dot{\Psi}$),
and identifying Misner's time $\Omega$ with the classical Hamiltonian time.
Thus, one is led to
%%%%%%%%%%%%%%%%%%%%%%%%%%
\begin{eqnarray}
\dot{q}\equiv\frac{dq}{d\Omega}&=&e^{-Q\Omega}p~\\   %p& ~\\
\dot{p}\equiv
\frac{dp}{d\Omega}&=&\kappa e^{(Q-4)\Omega}q~.
%\frac{Q^2}{4}+e^{-4\Omega}~.
\end{eqnarray}
%%%%%%%%%%%%%%%%%%%
These equations describe the canonical
motion for a classical EFRW point universe as derived from the
time-dependent Hamiltonian of the inverted oscillator type [9]
%%%%%%%%%%%%%%%%%%%%%%%%%%%%%%%%%%%%%%%%%%%%%%%%%%%%%%%%%%%%%%%%%%%%%%%%%
\begin{equation} \label{4}
H_{\rm cl}(\Omega)=
e^{-Q\Omega}\frac{p^2}{2}-\kappa e^{(Q-4)
\Omega}\frac{q^2}{2}~.
\end{equation}
%%%%%%%%%%%%%%%%%%%%%%%%%%%%%%%%%%%%%%%%%%%%%%%%%%%%%%%%%%%%%%%%%%%%%%%%%%
%\break{0.98in}
For this classical EFRW Hamiltonian the triplet of phase-space
functions $T_1=\frac{p^2}{2}$, $T_2=pq$,
%\break{0.35in}
and $T_3=\frac{q^2}{2}$ forms a dynamical Lie algebra (i.e.,
$H=\sum _{n}h_{n}(\Omega)T_{n}(p,q)$) which is closed with
respect to the Poisson bracket, or more exactly
$\{T_1,T_2\}=-2T_1$, $\{T_2,T_3\}=-2T_3$, $\{T_1,T_3\}=-T_2$. Using this
algebra $H_{\rm cl}$ reads
%%%%%%%%%%%%%%%%%%%%%%%%%%%%%%%%%%%%%%%%%%%%%%%%%%%%%%%%%%%%%%%%%%%%%%%%%%%
\begin{equation} \label{H}
H_{\rm cl}=e^{-Q\Omega}T_1-\kappa e^{(Q-4)\Omega}T_3~.
\end{equation}
%%%%%%%%%%%%%%%%%%%%%%%%%%%%%%%%%%%%%%%%%%%%%%%%%%%%%%%%%%%%%%%%%%%%%%%%
%\break{1.178in}
The Ermakov invariant ${\cal I}$ belongs to the dynamical algebra, i.e.,
one can write ${\cal I}=\sum _{r}\epsilon _{r}(\Omega)T_{r}$, and by means of
%%%%%%%%%%%%%
%\break{1.18in}
%%%%%%%%%%%%%%%%%
$\frac{\partial {\cal I}}{\partial \Omega}=-\{{\cal I},H\}$ one is led to the
following equations for the functions $\epsilon _{r}(\Omega)$
%%%%%%%%%%%%%%%%%%%%%%%%%
\begin{equation} \label{7}
\dot{\epsilon} _{r}+\sum _{n}\Bigg[\sum _{m}C_{nm}^{r}h_{m}
(\Omega)\Bigg]\epsilon _{n}=0~,
\end{equation}
%%%%%%%%%%%%%%%%%%%%%%%%
where $C_{nm}^{r}$ are the structure constants of the Lie algebra that
have been already given above. Thus, we get
%%%%%%%%%%%%%%
\begin{eqnarray} \nonumber
\dot{\epsilon} _1&=&-2e^{-Q\Omega}\epsilon _2 \\
\dot{\epsilon} _2&=&-\kappa e^{(Q-4)\Omega}\epsilon _1-
e^{-Q\Omega}\epsilon _3\\
\dot{\epsilon} _3&=&-2\kappa e^{(Q-4)\Omega}\epsilon _2~. \nonumber
\end{eqnarray}
%%%%%%%%%%%%%%
The solution of this system can be readily obtained by setting
$\epsilon _1=\rho ^2$ giving $\epsilon _2=-e^{Q\Omega}\rho
\dot{\rho}$ and
$\epsilon _3=e^{2Q\Omega}\dot{\rho} ^2 +
\frac{1}{\rho ^2}$, where $\rho$ is the solution of the Milne-Pinney (MP)
equation [10],
%%%%%%%%%%%%%%%%%%%%%%%%%%
$%\begin{equation} \label{9}
\ddot{\rho}+Q\dot{\rho}-\kappa e^{-4\Omega}\rho=\frac{e^{-2Q\Omega}}
{\rho ^3}$.
There is a well-defined prescription going back to Pinney's note in 1950
of writing $\rho$ as a
function of the particular solutions of the corresponding parametric oscillator
problem, i.e., the modified Bessel functions in the EFRW case.
In the formulas herein, we shall keep the symbol $\rho$ for this known
function.
%The method asks for a constant Wronskian ${\cal W}$ of the independent
%solutions of the gauge-transformed form of Eq.(1) (without the first derivative
%term). Since
%$W[K_{\alpha}(z),I_{\alpha}(z)]=1/z$, one gets a constant ${\cal W}=-2$,
%as it should be.
%only by fixing $Q=2$, that will be assumed from now on.
In terms of the function $\rho (\Omega)$ the Ermakov invariant reads [11]
%%%%%%%%%%%%%%
\begin{equation} \label{10}
{\cal I} _{\rm EFRW}=\frac{(\rho p-e^{Q\Omega}\dot{\rho}q)^2}{2}
+\frac{q^2}{2\rho ^2}=
\frac{e^{2Q\Omega}}{2}\left(
\rho \dot{\Psi} _{\alpha}-\dot{\rho} \Psi _{\alpha}\right)^2
+\frac{1}{2}\left(\frac{\Psi _{\alpha}}{\rho}
\right)^{2}~.
%e^{4\Omega /\kappa}\frac{(\rho\dot{\Psi _{1}}-\dot{\rho}\Psi _{1})^2}{2}+
%\frac{\Psi _{1} ^2}{2\rho ^2}~.
\end{equation}
%%%%%%%%%%%%%%

{\bf 2.}
Next, we calculate the time-dependent generating function allowing one to
pass to new canonical variables for which ${\cal I}$ is chosen to be the
new ``momentum"
%%%%%%%%%%%%%%%%%%%%%%
$S(q,P={\cal I},\vec{\epsilon}(\Omega))=\int ^{q}dq^{'}p(q^{'},
{\cal I},\vec{\epsilon}
(\Omega))$ leading to
%%%%%%%%%%%%%
\begin{equation}   \nonumber         %\label{12}
S(q,{\cal I},\vec{\epsilon}
(\Omega))=e^{Q\Omega}\frac{q^2}{2}\frac{\dot{\rho}}{\rho}+
{\cal I}{\rm arcsin}\Bigg[\frac{q}{\sqrt{2{\cal I}\rho ^2}}\Bigg]
         +\frac{q\sqrt{2{\cal I}\rho ^2-q^2}}{2\rho ^2}~, %+{\rm const}
\end{equation}
%%%%%%%%%%%%%%%%%%%%%%%%%%%
where we have put to zero the constant of integration.
Thus,
%%%%%%%%%%%%%%%%%%%%
$
\theta=\frac{\partial S}{\partial {\cal I}}={\rm arcsin}
\Big(\frac{q}{\sqrt{2{\cal I}\rho ^2}}\Big)$. Moreover, the canonical variables
are now $q=\rho \sqrt{2{\cal I}}\sin \theta$
and
%%%%%%%%%%%%%%
$p=\frac{\sqrt{2{\cal I}}}{\rho}\Big(\cos \theta+
e^{Q\Omega}\dot{\rho}\rho\sin \theta\Big)$.
%%%%%%%%%%%%%%%
The dynamical angle will be
%%%%%%%%%%%%%
$\Delta \theta ^{d}=
\int _{\Omega _{0}}^{\Omega}
\langle\frac{\partial H_{\rm{new}}}{\partial {\cal I}}\rangle
d\Omega ^{'}=
\int _{0}^{\Omega}[\frac{e^{-Q\Omega '}}{\rho ^2}-\frac{\rho ^2}{2}
\frac{d}{d\Omega ^{'}}\Big(\frac{\dot{\rho}}{\rho}\Big)]d\Omega ^{'}$,
whereas the geometrical angle reads
$\Delta \theta ^{g}=\frac{1}{2}\int _{\Omega _0}^{\Omega}
[\frac{d}{d\Omega ^{'}}
(e^{Q\Omega ^{'}}\dot{\rho}\rho)-2e^{Q\Omega ^{'}}
\dot{\rho}^2]
d\Omega ^{'}$.
%%%%%%%%%%%%%%%
Thus, the total change of angle is
%%%%%%%%%%%%%%%%%%%%%%
$\Delta \theta =\int _{\Omega _{0}}^{\Omega}\frac{e^{-Q\Omega ^{'}}}
{\rho ^2}
d\Omega ^{'}$.
%%%%%%%%%%%%%%%%%%%%%%%%
On the Misner time axis, going to $-\infty$ means going to the origin of the
universe, whereas $\Omega _{0}=0$ means the present epoch. Using
these cosmological limits we obtain {\em the
interesting result} that the total change
of angle $\Delta \theta$ during the cosmological evolution in $\Omega$ time
can be written up to a sign as the Laplace transform of parameter $Q$
of the inverse square of the MP function, $\Delta \theta=
-L_{1/\rho ^{2}}(Q)$.
%For periodic parameters $\vec{\mu}(\Omega)$, with all the components of
%the same period $T$, the
%geometric angle is known as the (nonadiabatic) Hannay angle \cite{book},
%but we are not in such a situation here.
%and in terms of $\rho$ is given by
%%%%%%%%%%
%\begin{equation}  \label{17b}
%\Delta \theta ^{g}_{H}=-\oint _{C}\dot{\rho}d\rho~.
%\end{equation}
%%%%%%%%%%%%%%%

%%%%%%%%%%%%%%%%%%%%%%%%%%%%%%%%%%%%%%%%%%%%%%%%%%%%%%%%%%%%%%%%%%%%%%%%%%
{\bf 3}.
We now sketch the minisuperspace Taub model
%which has been studied in some
%detail by Ryan and collaborators \cite{ry} who found tthat it is separable.
for which the WDW equation reads
$$
\frac{\partial ^2\Psi}{\partial \Omega ^2}-
\frac{\partial ^2\Psi}{\partial \beta ^2}+Q\frac{\partial \Psi}
{\partial \Omega}+e^{-4\Omega}V(\beta)\Psi=0~,
\eqno(10)
$$
where $V(\beta)=\frac{1}{3}(e^{-8\beta}-4e^{-2\beta})$.
%Again the first
%derivative term can be eliminated and the resulting
This equation can be separated in the variables
$x_1=-4\Omega-8\beta$ and $x_2=-4\Omega -2\beta$.
Thus, one gets the following two independent 1D problems for which the Ermakov
procedure can be repeated along the lines of the EFRW case
%The two 1D equations are as follows
$$
\frac{d ^2 \Psi _{T1}}{d x_1^2}+\frac{Q}{12}\frac{d\Psi _{T1}}{dx_1}
+\left(\frac{\omega ^2}{4}-\frac{1}{144}e^{x_1}\right)
\Psi _{T1}=0  %\frac{d ^2 \Phi _{T1}}{d x_1^2}+W_1(x_1)\Phi _{T1}=0
%\frac{\omega ^2}{4}u_1
\eqno(11)
$$
and
$$
\frac{d^2 \Psi _{T2}}{d x_2^2}-\frac{Q}{3}\frac{d\Psi _{T2}}{dx_2}
+\left(\omega ^2-\frac{1}{9}e^{x_2}\right)\Psi _{T2}=0~.
%\frac{d^2 \Psi _{T2}}{d x_2^2}+W_2(x_2)\Psi _{T2}=0~,
\eqno(12)
$$
%where we have already multiplied both sides in Eqs. (3) and (4) by
%$-1$ in order to get standard Schr\"odinger equations.
The quantity $\omega /2$ is the separation constant.
%which physically is related to the wavenumber of a positive energy level in a
%Schroedinger interpretation.
%Mart\'{\i}nez and Ryan have considered a wavepacket solution made of
%wavefunctions $\Psi$ having
%the form of a product of modified Bessel functions of imaginary order.
The solutions are $\Psi _{T1}\equiv \Psi _{T\alpha_{1}}
=e^{(-Q/24)x_1}Z_{i\alpha _1}(ie^{x_1/2}/6)$ and
$\Psi _{T2}\equiv \Psi _{T\alpha _{2}}
=e^{(Q/6)x_2}Z_{i\alpha _2}(i2e^{x_2/2}/3)$,
%+K_{i\alpha _2}(e^{x_2/2}/3)]$,
respectively, where $\alpha _1=
\sqrt{\omega ^2-(Q/12)^2}$ and $\alpha _2=\sqrt{4\omega ^2-(Q/3)^2}$.

{\bf 4}.
A more realistic case is provided by the minimally
coupled FRW-massive-scalar-field minisuperspace model. The Ermakov approach,
which differs from the previous one,
will be studied in detail elsewhere. The WDW equation reads
$$
[\partial ^2_{\Omega}+Q\partial _{\Omega}-\partial ^{2}_{\phi}-\kappa
e^{-4\Omega}+m^2e^{-6\Omega}\phi ^2]\Psi (\Omega ,\phi)=0~,
\eqno(13)
$$
and can be written
as a two-component Schroedinger equation (see e.g., [12]). This allows one
to think of cosmological squeezed states based on the Ermakov approach
[13,14]. For this one makes use of the
factorization of the Ermakov invariant ${\cal I}
=\hbar(bb^{\dagger}+\frac{1}{2})$,
where
%%%%%%%%%%%%%
%\begin{eqnarray} %\nonumber
$
b          =(2\hbar )^{-1/2}
  [\frac{q}{\rho}+i(\rho p-e^{Q_{c}\Omega}\dot{\rho}q)]
$
and
$
b^{\dagger}=
(2\hbar )^{-1/2}[\frac{q}{\rho}-i(\rho p-e^{Q_{c}\Omega}
\dot{\rho}q)]
$.
%\end{eqnarray}
%%%%%%%%%%%%%%%%%%%%%%
$Q_{c}$ is a fixed HH factor ordering parameter.
Let us now consider a reference Misner-time-independent oscillator with
the Misner frequency $\omega _{0}$ corresponds to an arbitrary
epoch $\Omega _{0}$ for which one can write the common factorizing operators
%%%%%%%%%%%%%%%%%%
$a          =(2\hbar \omega _{0})^{-1/2}[\omega _{0}q+ip],\quad
a^{\dagger}=(2\hbar \omega _{0})^{-1/2}[\omega _{0}q-ip]
$.
%%%%%%%%%%%%%%%%%%%
The connection between the $a$ and $b$ pairs is given by
%%%%%%%%%%%%%
$b(\Omega)          =\mu(\Omega)a+\nu(\Omega)a^{\dagger}$ and
$b^{\dagger}(\Omega)=\mu ^{*}(\Omega)a^{\dagger}+\nu ^{*}
(\Omega)a^{\dagger}~, %\nonumber
$
%%%%%%%%%%%%%%%%%%
where
%%%%%%%%%
$\mu(\Omega)=(4\omega _{0})^{-1/2}[\rho ^{-1}-ie^{Q_{c}\Omega}
\dot{\rho}+\omega _{0}\rho]$
%%%%%%%%%%%%%%%%
and
%%%%%%%%%
$\nu(\Omega)=(4\omega _{0})^{-1/2}[\rho ^{-1}-ie^{Q_{c}\Omega}
\dot{\rho}-\omega _{0}\rho]$ fulfill the well-known
relationship $|\mu(\Omega)|^{2}-|\nu(\Omega)|^{2}=1$.
The corresponding uncertainties are known to be
$(\Delta q)^{2}=\frac{\hbar}{2\omega _{0}}|\mu - \nu|^{2}$,
$(\Delta p)^{2}=\frac{\hbar \omega _{0}}{2}|\mu + \nu|^{2}$, and
$(\Delta q)(\Delta p)= \frac{\hbar}{2}|\mu +\nu||\mu -\nu|$ showing that
in general the Ermakov squeezed states are not minimum uncertainty
states [14].

%In conclusion, a simple cosmological application of the Ermakov procedure
%has been presented here based on a
%classical point particle representation of pure gravity WDW equations.

{\bf 5}.
As was first noticed by K.R. Symon [15], the Ermakov invariant
is equivalent to the Courant-Snyder one in accelerator physics, which defines
the admittance of the accelerating device. This
allows in a certain sense a beam physics approach to cosmological evolution.
The point is that under the assumption of no coupling
between the radial and the vertical betatron oscillations, the latter ones
are described by the Hill
equation $z^{''}+n(s)\kappa _{o}^2z=0$, where $n$ is the
magnetic field index, and $\kappa _{o}$ is the curvature of the orbit which is
parametrized by $s$ that may be considered as a counterpart of $\Omega$.
The solutions can be written as $z_{\pm}={\rm w}(s)e^{\pm i\psi(s)}$ and
only a real $\psi$ leads to bounded oscillations. The
amplitude ${\rm w}(s)$ satisfies
a MP equation and moreover $\psi={\rm w}^{-2}$. However,
in ``quantum" cosmology, $\psi$ is a pure imaginary action functional
leading to instabilities in the $\Psi$ solutions. In other words,
while in accelerators we are interested in stable periodic solutions,
in ``quantum"
cosmology there are the unstable parametric solutions that come into play.

\bigskip
%%%%%%%%%%%%%%%%%%%%%%%%%%%%%%%%%%%%%%%%%%%%%%%%%%%%%%%%%%%%%%%%%%%%%
%\nonumsection{Acknowledgment}

\noindent
This work was partially supported by the CONACyT Project 458100-5-25844E.
We wish to thank M.A. Reyes for very helpful discussions.
%and 3898P-E9608.

%\begin{thebibliography} {000}

\bigskip
\bigskip

\centerline{REFERENCES}

\bigskip

%\bibitem[*]{byline} E-mail: rosu@ifug.ugto.mx

%\bibitem{erm}
\noindent
[1] V.P. Ermakov, Univ. Izv. Kiev. Ser. III {\bf 9} (1880) 1;
         For a recent review, see R.S. Kaushal,
         Int. J. Theor. Phys. {\bf 37} (1998) 1793

%\bibitem{work}
\noindent
[2] See for example,
         S. Cotsakis, R.L. Lemmer, P.G.L. Leach, Phys. Rev.
         {\bf D 57} (1998) 4691; %Adiabatic inv's and mixmaster catastrophes
         S.P. Kim, Class. Quantum Grav. {\bf 13} (1996) 1377

%\bibitem{book}
\noindent
[3] M.V. Berry, Proc. R. Soc. London {\bf A 392} (1984) 45;
         J.H. Hannay, J. Phys. {\bf A 18} (1985) 221;
         A. Shapere, F. Wilczek, Geometric Phases in Physics.
         World Scientific, Singapore 1989; J. Anandan, J. Christian,
         K. Wanelik, Am. J. Phys. {\bf 65} (1997) 180

%\bibitem{dutta}
\noindent
[4] D.P. Dutta, Phys. Rev. {\bf D 48} (1993) 5746;
         Mod. Phys. Lett.
         {\bf A 8} (1993) 191, 601
         % ``Semiclassical BR and Berry's phase"

%\bibitem{rosu}
\noindent
[5] See for example, V. Moncrief, M.P. Ryan, Phys. Rev. {\bf D 44} (1991)
        2375; O. Obreg\'on, J. Socorro, Int. J. Theor. Phys. {\bf 35} (1996)
        1381; H.C. Rosu, Mod. Phys. Lett. {\bf A 13} (1998) 227

\noindent
[6] For $Q=0,\;\kappa =1$ see H. Rosu, J. Socorro, Nuovo Cim. {\bf B 113}
                              (1997) 683

%\bibitem{hh}
\noindent
[7] J. Hartle, S.W. Hawking, Phys. Rev. {\bf D 28} (1983) 2960

%\bibitem{mi}
\noindent
[8]  C.W. Misner, Phys. Rev. Lett. {\bf 22} (1969) 1071;
         Phys. Rev. {\bf 186} (1969) 1319, 1328

%\bibitem{clas}
%         For a recent paper see B. Mielnik, M.A. Reyes,
%         J. Phys. {\bf A 29} (1996) 6009

%\bibitem{inv}
\noindent
[9]
         See for example, S. Baskoutas and A. Jannussis,
         Nuovo Cim. {\bf B 107} (1992) 255
         %; A thorough discussion of the
         %normal (not inverted) oscillator (pendulum) with time dependence in
         %both mass and frequency may be found in P.G.L. Leach, SIAM J. Appl.
         %Math. {\bf 34} (1978) 496

%\bibitem{rr1}
%      H.C. Rosu and J.L. Romero, ``Ermakov approach for the one-dimensional
%         Helmholtz Hamiltonian", quant-ph/9707044.

%\bibitem{2}
%         F. Delgado, B. Mielnik, and M. Reyes, ``Helmholtz spectra and
%         particle optics", preprint, 1997.

%\bibitem{3}
%         H. Pr\"ufer, Math. Annalen {\bf 95}, 409 (1926);

%\bibitem{mp}
\noindent
[10]  W.E. Milne, Phys. Rev. {\bf 35} (1930) 863;
         E. Pinney, Proc. Am. Math. Soc. {\bf 1} (1950) 681

%\bibitem{l}
\noindent
[11] H.R. Lewis Jr., J. Math. Phys. {\bf 9} (1968) 1976

\noindent
[12] A. Mostafazadeh, J. Math. Phys. {\bf 39} (1998) 4499

%\bibitem{lr}
%noindent
%11] H.R. Lewis, W.B. Riesenfeld, J. Math. Phys. {\bf 10}
%        (1969) 1458

%%%%%%%%%%%%%%%%%%%%%%%%%%%%%%%%%%%%%%%%%%%%%%%%%%%%%%%%%%%%%%
%\bibitem{mor}
%         D.A. Morales, J. Phys. A {\bf 21}, L889 (1988).

%\bibitem{m}
%         M. Maamache, Phys. Rev. A {\bf 52}, 936 (1995).

%\bibitem{hk}
%         K. Hayata and M. Koshiba, Opt. Lett. {\bf 20}, 1131 (1995).
%%%%%%%%%%%%%%%%%%%%%%%%%%%%%%%%%%%%%%%%%%%%%%%%%%%%%%%%%%%%%%%%%%%%

%\bibitem{sq1}
\noindent
[13] J.G. Hartley, J.R. Ray, Phys. Rev. {\bf D 25}
         (1982) 382

%\bibitem{sq2}
\noindent
[14] I.A. Pedrosa, Phys. Rev. {\bf D 36} (1987) 1279;
         I.A. Pedrosa, V.B. Bezerra, Mod. Phys. Lett. {\bf A 12}
         (1997) 1111

%\bibitem{cs}
\noindent
[15]  Second footnote in [11];
         E.D. Courant, H.S Snyder,
         Ann. Phys. (N.Y.) {\bf 3} (1958) 1 (formula 3.22).
         %See formula 3.22 therein.

%\end{thebibliography}

%$The length of your paper, including references and figures, should not
%exceed 4 pages.  Manuscripts longer than 4 pages will not be printed.
%For references, tables, figures, use the Physical Review style given in
%RevTex.  Please send two hard copies of your manuscript to
%\medskip
%\begin{center}
%Dr. Daesoo Han \\
%NASA -- Goddard Space Flight Center, Code 935 \\
%Greenbelt, Maryland 20771, U.S.A.
%\medskip
%\end{center}
%The text page of your hard copy should be 21.5 cm (8.5 in) wide
%and 27.9 cm (11 in) high, with top, bottom, left and right margins
%of 2.5 cm (1 inch).  The text size is therefore 16.5 by 22.9 cm.
%The RevTex gives this text size.  Do not put page numbers on text
%pages.  Put the number with pencil on the back of each page.
%Please enclose also the diskette containing your manuscript, or send
%the file by e-mail to $<han@trmm.gsfc.nasa.gov>$.
%\medskip
%\begin{center}
% The deadline for submission is 31 July 1999.
%\end{center}
%\medskip

%If you cannot do RevTex or LaTex, type your manuscript within the
%text page of  21.5 cm (8.5 in) wide and 27.9 cm (11 in) high, with
%top, bottom, left and right margins of 2.5 cm (1 inch).  The text
%size is therefore 16.5 by 22.9 cm.  Follow this RevTex-based format
%as closely as possible.

\end{document}